\documentclass[%
 reprint,
 superscriptaddress,
 preprintnumbers,
 nofootinbib,
 amsmath,amssymb,
 aps,
 prd,
floatfix,
]{revtex4-2}

\usepackage{graphicx}
\usepackage{dcolumn}
\usepackage{bm}

\usepackage{physics} 
\usepackage{siunitx} 
\usepackage{enumerate} 
\usepackage{tabularx}
\usepackage{amssymb}
\usepackage{amsmath,mathtools,mathrsfs,slashed}
\usepackage{subcaption}
\usepackage{xcolor}

\newcommand{\gE}{\Tilde{g}}
\newcommand{\RE}{\Tilde{R}}
\newcommand{\DE}{\Tilde{\nabla}}
\newcommand{\BE}{\Tilde{\square}}
\newcommand{\KE}{\Tilde{K}}

\renewcommand{\i}{{\mathrm{i}}}

\newcommand{\STa}{\alpha}
\newcommand{\STb}{\beta}
\newcommand{\STc}{\tau}
\newcommand{\STd}{\rho}

\usepackage{feynmf}

\newcommand{\higgsdilaton}{\cite{Ferreira:2016vsc,Casas:2017wjh,Ferreira:2018qss,Shaposhnikov:2008xb,Buchmuller:1988cj,Shaposhnikov:2008xi,Blas:2011ac,Garcia-Bellido:2011kqb,Garcia-Bellido:2012npk,Bezrukov:2012hx,Henz:2013oxa,Rubio:2014wta,Karananas:2016grc,Bars:2013yba,Bezrukov:2014ipa,Fujii:1974bq,Shaposhnikov:2022dou,Karananas:2022byw,Karananas:2021gco,Shaposhnikov:2018jag,Ghilencea:2022lcl,Rubio:2020zht}}
\newcommand{\frameineq}{\cite{Capozziello:1996xg,Capozziello:2010sc,Nojiri:2000ja,Kamenshchik:2014waa,Postma:2014vaa,Banerjee:2016lco,Pandey:2016unk,Alvarez:2001qj,Karam:2017zno,Pandey:2016jmv,Bounakis:2017fkv,Karam:2018squ,Faraoni:1999hp,Briscese:2006xu,Capozziello:2006dj,Bahamonde:2016wmz,Bahamonde:2017kbs,Frion:2018oij,Chiba:2013mha,Domenech:2015qoa,Steinwachs:2013tr,Jarv:2016sow,Herrero-Valea:2016jzz}}

\begin{document}


\title{Fifth forces and frame invariance}

\author{Jamie Bamber}
\email{james.bamber@physics.ox.ac.uk}
\affiliation{Astrophysics, University of Oxford, Denys Wilkinson Building, Keble Road, Oxford OX1 3RH, United Kingdom}

\date{Received \today; published -- 00, 0000}

\begin{abstract}
I discuss how one can apply the covariant formalism developed by Vilkovisky and DeWitt to obtain frame invariant fifth force calculations for scalar-tensor theories. Fifth forces are severely constrained by astrophysical measurements. It was shown previously that for scale-invariant Higgs-dilaton gravity, in a particular choice of Jordan frame, the dilaton fifth force is dramatically suppressed, evading the observational constraints. Using a geometric approach I extend this result to all frames, and show that the usual dichotomy of ``Jordan frame" versus ``Einstein frame" is better understood as a continuum of frames: submanifold slices of a more general field space. 
\end{abstract}

\maketitle

\section{Introduction}

Since Einstein formulated his theory of General Relativity over a century ago \cite{EinsteinAlbert1987Tcpo} there has been much theoretical interest in the possibility that it is merely an approximation to a more general theory of gravity. One of the most popular classes of theories of modified gravity are the so-called ``scalar-tensor" theories \cite{Kobayashi:2019hrl}, where the Einstein-Hilbert action \footnote{note that throughout we assume a mostly plus signature $(-,+,+,+)$.}
\begin{equation}
    S = \int \dd^4 x \sqrt{-g}\left[\frac{M^2_{\textup{Pl}}}{2}R + L_{\textup{m}} \right],
\end{equation}
is modified by the addition of one or more scalar fields. $L_{\textup{m}}$ is the matter part of the Lagrangian\footnote{There is sometimes ambiguity as to whether the Lagrangian is defined with or without the $\sqrt{-g}$ term. I will be using curly $\mathcal{L}$ to refer to Lagrangian including the $\sqrt{-g}$ metric factor, and upright $L$ when not including the $\sqrt{-g}$.}. Instead of a fixed Planck mass $M_{\textup{Pl}}$ (or alternatively a fixed Newton's constant $G$) we introduce a non-trivial coupling to $R$, giving an action of the form
\begin{equation}
    S = \int \dd^4 x \sqrt{-g}\Bigg[F(\boldsymbol{\varphi})R - \frac{1}{2}\partial_{\mu}\boldsymbol{\varphi}\cdot \partial^{\mu}\boldsymbol{\varphi} - W(\boldsymbol{\varphi}) + L_{\textup{m}}\Bigg], \label{eq:Jordan}
\end{equation}
where the effective Planck mass is now a function of the scalar field(s) $\boldsymbol{\varphi} = \{\phi_i\}$. The first and arguably simplest theory of this type is that of Brans-Dicke from 1961 \cite{Brans:1961sx} where $F(\phi)=-\frac{\alpha}{12}\phi^2, \, W=0$ for a single scalar field $\phi$ and constant $\alpha$. A modern formulation which encompasses all possible scalar-tensor theories with second order equations of motion was given by Horndeski \cite{Horndeski:1974wa,Kobayashi:2019hrl}. 

One feature of these theories is that they can be expressed in different guises or ``frames" via field redefinitions. Equation \eqref{eq:Jordan} describes a ``Jordan" frame if $F(\boldsymbol{\varphi})$ depends on $\boldsymbol{\varphi}$. With a suitable Weyl transformation $g_{\mu\nu} \rightarrow \Omega^2(\boldsymbol{\varphi}) \Tilde{g}_{\mu\nu}$ we can obtain a new action $S = \int \dd^4 x \sqrt{-\Tilde{g}}\left[\frac{M^2}{2}\Tilde{R} + \dots + L_{\textup{m}}(\Omega(\boldsymbol{\varphi}),\Tilde{g}_{\mu\nu},\textup{matter})\right]$, where the gravity sector is now as in GR, but the matter sector picks up additional couplings to $\boldsymbol{\varphi}$. This is termed the ``Einstein frame". 

Despite the theoretical attractions (Paul Dirac argued for a dynamical $G$ on the basis of his large number hypothesis) generic scalar-tensor theories are severely constrained by solar system and lab observations. This is because the introduction of an additional field with a non-trivial coupling to gravity or matter can in general mediate long-range ``fifth forces" \cite{Brans:1961sx,Damour:1992we,Poisson:2014}\footnote{So-called because they act in addition to the standard four fundamental forces of nature.}. The exchange of a new particle of mass $m$ coupling to matter gives a Yukawa \cite{Yukawa:1935xg} potential
\begin{equation}
    V_{\textup{fifth}}(r) = - \frac{\epsilon^2 M_1 M_2}{4\pi r} e^{-mr},
\end{equation}
for coupling $\epsilon$ and masses $M_1,M_2$. For small enough $m$ ($m = 0$ for Brans-Dicke) this can be probed via solar system tests, which put extremely tight bounds on $\epsilon$ \cite{Gonzalez:2020vzl,Bertotti:2003rm,Freire:2012mg}. In other words if a theory predicts a significant long-range fifth force, like standard Brans-Dicke, it is probably ruled out. 

One particularly interesting type of scalar-tensor theory is one that is scale-invariant (including ``Higgs-dilaton" theories where one of the scalar fields is a non-minimally coupled SM Higgs boson) \higgsdilaton, where the action, including the matter sector, has a global Weyl symmetry \cite{Weyl:1919rzm:} such that there is no a-priori lengthscale. Instead the symmetry is broken dynamically as the scalar field(s) tend towards fixed equilibrium values under the influence of an expanding cosmology. This has been proposed as one element of a solution to the so-called hierarchy problem \cite{Shaposhnikov:2018jag,Foot:2007iy,Shaposhnikov:2008xi}, and the phenomenological implications of such a theory have also generated substantial interest \cite{Kamada:2012se,Greenwood:2012aj,Rinaldi:2015uvu,Barrie:2016rnv,Ghilencea:2021jjl,Ghilencea:2021lpa,Aoki:2021skm,Ferreira:2019ywk}. There is a massless Goldstone boson associated with the spontaneously broken symmetry, termed the ``dilaton", $\sigma$, and as such we might be worried about fifth-force constraints. However, Ferreira, Hill \& Ross (2016) \cite{Ferreira:2016kxi} show that, in a particular choice of Jordan frame, the dilaton completely decouples from the matter sector, and therefore contributes no fifth force. 

This leads to the question: does this result apply in other choices of frame? Indeed, to what extent are generic scalar-tensor theories of this type really physically equivalent in different frames? While on a classical level one should not expect a redefinition of variables to change the physics or physical results, once you include quantum corrections this becomes no longer obvious (this is sometimes called the ``cosmological frame problem" \cite{Karamitsos:2018lur}). Copeland et al. \cite{Copeland:2021qby} and Burrage et al. \cite{Burrage:2018dvt} explicitly calculated the fifth forces for a three-scalar-field toy model, which becomes a Higgs-dilaton theory for a certain choice of parameters, in first the Einstein frame \cite{Copeland:2021qby} and a Jordan frame \cite{Burrage:2018dvt}, and showed that at lowest perturbative order the results are the same. There has also been extensive work examining the general question of frame (in)equivalence from numerous points of view, mostly focused on cosmological applications \frameineq.

In particular, Falls \& Herrero-Valea \cite{Falls:2018olk,Herrero-Valea:2020qgj,Herrero-Valea:2016jzz} and Finn et al. \cite{Finn:2021jdz,Finn:2022rlo} developed a formalism to characterise exactly how the quantum effective action must transform non-trivially between frames. Finn et al. \cite{Finn:2021jdz,Finn:2022rlo} adopts the covariant approach \cite{Burns:2016ric,Moss:2014nya,Cohen:2022uuw}, pioneered by Vilkovisky and DeWitt \cite{DeWitt:1985sg,DeWittBryceS1991Ser,Vilkovisky:1984st}, whereby frame transformations are described in terms of a coordinate changes on a field-space manifold, and constructs a fully covariant quantum effective action, extending the Vilkovisky-DeWitt unique effective action to theories with fermion fields.

In this paper I show how this formalism, and the covariant geometric approach, can be applied to the problem of computing fifth forces, and to scale invariant scalar-tensor theories in particular. I extend the geometric approach to show how choices of frame can be characterised in a geometric manner: as choices of submanifold in a higher dimensional general field space. Frame invariance becomes manifest, and we see how the choice of frame is better thought of not as a dichotomy between ``Jordan" and ``Einstein", but as a continuum one can smoothly traverse. We also see how scale-invariant scalar-tensor gravity evades fifth force constraints in all possible frames.

The structure of this paper is as follows. Section \ref{sec:bg} lays out the background theory. Section \ref{sec:geom} describes the new geometric approach to frame fixing, and in section \ref{sec:fith_force} I apply it to calculations of fifth forces. I focus on the scale invariant theory in section \ref{sec:scale_inv}, and briefly discuss one-loop corrections from the choice of \textit{physical} spacetime in section \ref{higher_order}. Finally I conclude with a discussion of my results and future directions. 


\section{Background}
\label{sec:bg}

\subsection{The dilaton and scale invariant gravity}
\label{sec:dilaton}

Under the Weyl transformation $g_{\mu\nu} = \Omega^2 \Tilde{g}_{\mu\nu}$ the Jordan frame action \eqref{eq:Jordan} for some integer number of scalar fields $\boldsymbol{\varphi} = \{\phi_i\}$ becomes
\begin{equation}
\begin{split}
    S =& \int \dd^4 x \sqrt{-\gE}\bigg[F(\boldsymbol{\varphi}) \Omega^2 \left(\RE - 6(\DE \ln \Omega)^2 - 6\BE \ln \Omega\right) \\
    &- \Omega^2\frac{1}{2}\sum_i \partial_{\mu} \phi_i \partial^{\mu} \phi_i - \Omega^4 W(\boldsymbol{\varphi}) + \Omega^4L_{\textup{m}}\bigg],
\end{split}
\end{equation}
where $(\DE v)^2 := (\DE_{\mu} v)(\DE^{\mu} v)$. Let 
\begin{alignat}{11}
    &\Omega &&= \exp(\sigma), &&\Tilde{F}(\sigma,\Tilde{\boldsymbol{\varphi}}) &&= \Omega^2F(\boldsymbol{\varphi}),   &&\Tilde{\phi}_i &&= \Omega \phi_i, \notag\\
    &\KE &&= \frac{1}{2}\sum_i \Tilde{\phi}^2 + 6\Tilde{F},  \quad&&\Tilde{W}(\sigma,\Tilde{\boldsymbol{\varphi}}) &&= \Omega^4 W(\boldsymbol{\varphi}), \quad &&\Tilde{L}_{\textup{m}} &&= \Omega^4L_{\textup{m}}, \label{trans_rules}
\end{alignat}
Then we obtain
\begin{equation}
\begin{split}
    S =& \int \dd^4 x \sqrt{-\gE}\Big[\Tilde{F}(\sigma,\Tilde{\boldsymbol{\varphi}}) \left(\RE - 6(\DE \sigma)^2 - 6\BE \sigma\right) \\
    &- \frac{1}{2}\sum_i {\Tilde{\phi}_i}^2(\DE \sigma)^2 
    + (\DE_{\mu} \sigma)\sum_i \Tilde{\phi}_i\DE^{\mu}\Tilde{\phi}_i \\
    &- \frac{1}{2}\sum_i \partial_{\mu} \Tilde{\phi}_i \partial^{\mu} \Tilde{\phi}_i - \Tilde{W}(\sigma,\Tilde{\boldsymbol{\varphi}}) +\Tilde{L}_{\textup{m}}\Big]. \label{master_eq}
\end{split}
\end{equation}
Integrating by parts gives
\begin{equation}
\begin{split}
    S =& \int \dd^4 x \sqrt{-\gE}\Big[\Tilde{F}(\sigma,\Tilde{\boldsymbol{\varphi}})\RE -\KE(\sigma,\Tilde{\boldsymbol{\varphi}})(\DE \sigma)^2 \\
    &+ \DE_{\mu} \sigma\DE^{\mu}\KE(\sigma,\Tilde{\boldsymbol{\varphi}}) - \frac{1}{2}\sum_i \partial_{\mu} \Tilde{\phi}_i \partial^{\mu} \Tilde{\phi}_i - \Tilde{W}(\sigma,\Tilde{\boldsymbol{\varphi}}) + \Tilde{L}_{\textup{m}}\Big].
\end{split}
\end{equation}
The Euler-Lagrange equation for the dilaton $\sigma$ gives
\begin{equation}
\begin{split}
    \Tilde{\square}\Tilde{K}-2(\Tilde{K}-\partial_{\sigma}\Tilde{K})\Tilde{\square}\sigma-2\Tilde{\nabla}_{\mu}&(\Tilde{K}-\partial_{\sigma}\Tilde{K})\Tilde{\nabla}^{\mu}\sigma = \\
    &= - \partial_{\sigma}\Tilde{W} - \Tilde{R}\partial_{\sigma}\Tilde{F},
\end{split}
\end{equation}
where $\partial_{\sigma}:=\pdv{\sigma}$. We can see from the transformation rules \eqref{trans_rules} that if we choose $F(\boldsymbol{\varphi})$ to be quadratic in $\phi_i$, the potential $W(\boldsymbol{\varphi})$ to be quartic in $\phi_i$, and the $L_{\textup{m}}$ to also rescale appropriately, then the action becomes \textit{scale invariant} and $\partial_{\sigma}\Tilde{F}=\partial_{\sigma}\Tilde{W}=0$. The dilaton is then massless, and appears in the action only via its derivatives. 
\begin{equation}
\begin{split}
    S =& \int \dd^4 x \sqrt{-\gE}\Big[\Tilde{F}(\Tilde{\boldsymbol{\varphi}})\RE -\KE(\Tilde{\boldsymbol{\varphi}})(\DE \sigma)^2 \\
    &+ \DE_{\mu} \sigma\DE^{\mu}\KE(\Tilde{\boldsymbol{\varphi}}) - \frac{1}{2}\sum_i \partial_{\mu} \Tilde{\phi}_i \partial^{\mu} \Tilde{\phi}_i - \Tilde{W}(\Tilde{\boldsymbol{\varphi}}) + \Tilde{L}_{\textup{m}}\Big].
\end{split}
\end{equation}
If we then choose $\Omega$ such that $\Tilde{K} = const.$ we obtain the particular Jordan frame described in \cite{Ferreira:2016kxi} and we see that the dilaton completely decouples from the other scalar fields and the other matter terms in $\Tilde{L}_{\textup{m}}$. The equation of motion for the dilaton reduces to a wave equation $\Tilde{\square}\sigma = 0$, and it can be set to zero. As a result there are no fifth forces from the dilaton\footnote{There is still a coupling between gravity and the dilaton via its contribution to the stress energy tensor $T_{\mu\nu}$, and thus sourcing curvature according to standard General Relativity, however this contribution also vanishes for $\sigma = 0$.}.

\subsection{The covariant formalism}

\label{sec:cov}

Start by considering the path integral 
\begin{equation}
    Z[J] = \int [\mathcal{D}^N \Phi] e^{-S[\boldsymbol{\varphi}] - J_a \Phi^a},
\end{equation}
where $[\mathcal{D}^N \Phi]$ is an appropriate measure over the function space for fields $\boldsymbol{\Phi} = \{\Phi^i\}$, $J_a$ is a source term. I use $i,j \dots$ indices to denote field species and $a,b \dots$ to denote DeWitt indices spanning both field species and position or momentum \cite{Falls:2018olk}. A frame transformation is a field reparameterisation $\Phi^i \rightarrow \Tilde{\Phi}^i(\boldsymbol{\Phi})$. 
In the covariant formalism we describe this as a transformation of coordinates on a ``configuration space" or ``field space" manifold \cite{Finn:2022rlo}. This has an associated line element $\dd s^2 = C_{ab} \dd \Phi^a \dd \Phi^b$
where $a,b$. We would like our path integral, action and measure to be frame/reparameterisation invariant, which leads us to define
\begin{equation}
    [\mathcal{D}^N \Phi] = V^{-1}_{\textup{gauge}}\sqrt{\textup{det}(C_{ab})}\Pi_a \frac{\dd \Phi^a}{\sqrt{2\pi}},
\end{equation}
where $V_{\textup{gauge}} = \int \Pi_a \frac{\dd \xi^a}{\sqrt{2\pi}} \sqrt{\textup{det}(\sigma_{ab}(\boldsymbol{\Phi}))}$ accounts for the volume of the gauge group, where $\dd \xi^a$ are the generators of the Lie algebra and $\sigma_{ab}$ is another metric. For the moment I assume all the fields are bosonic, however this formalism has also been extended to include fermionic fields \cite{Finn:2022rlo} as I discuss later. To ensure diffeomorphism invariance of the free action, the preferred field space metric for four dimensions is \cite{Finn:2021jdz,Finn:2022rlo}
\begin{equation}
    C_{ab} = \frac{\bar{g}_{\mu\nu}}{4}\frac{\delta^2 S}{\delta(\partial_{\mu}\Phi^a)\delta(\partial_{\mu}\Phi^b)} = C_{ij} \bar{\delta}^{(4)}(x_a - x_b),
\end{equation}
where the delta function enforces locality. We assume $\sigma_{ab} = \sigma_{\mu\nu}\bar{\delta}^{(4)}(x_a - x_b)$ is also ultra-local \cite{Falls:2018olk}. The $\bar{g}_{\mu\nu}$ is the physical, or preferred, spacetime metric which satisfies dimensionless line element $\dd \bar{s}^2 = \bar{g}_{\mu\nu}\dd x^{\mu} \dd x^{\nu}.$ Defining $\bar{g}_{\mu\nu}$ is important to overcome the ambiguity between the physical space time metric and the gravity quantum field $g_{\mu\nu}$ \cite{Falls:2018olk,Finn:2021jdz,Finn:2022rlo}. The two are related by
\begin{align}
    \bar{g}_{\mu\nu} =& \;l^{-2}(\boldsymbol{\Phi})g_{\mu\nu}, \\
    =& \;e^{-2\sigma_{\textup{phys}}}g_{\mu\nu}, \\
    =& \;e^{2(\sigma-\sigma_{\textup{phys}})}\Tilde{g}_{\mu\nu},
\end{align}
where $l(\boldsymbol{\Phi})$ is an effective Planck length \cite{Finn:2022rlo}, and the functional derivatives are defined using the barred metric, and $\bar{\delta}^{(4)}(x)$ is defined such that 
\mbox{$\int \dd^4 x \sqrt{\bar{g}} \bar{\delta}^{(4)}(x) = 1$}.
For an action with kinetic term 
\begin{equation}
    S = \int \dd^4 x \left[-N_{ij}\Tilde{g}^{\mu\nu}\partial_{\mu}\Phi^i\partial_{\nu}\Phi^j + \dots\right],
\end{equation}
(summation implied) the associated field space metric is 
\begin{equation}
    C_{ij} = e^{2(\sigma-\sigma_{\textup{phys}})}N_{ij}.
\end{equation}
In theories without gravity we can take $\sigma_{\textup{phys}} = \sigma$ and canonically normalise the kinetic term so that $N_{ij} = const.$, allowing us to neglect the field space metric entirely as it only contributes a overall constant to the path integral. However, in theories with gravity the choice of $\sigma_{\textup{phys}}$ is important, and $C_{ij}$ can have non-trivial dependence on the fields $\Phi^i$. In order to obtain perturbative scattering amplitudes we expand about a flat Minkowski background
$g_{\mu\nu} \approx \eta_{\mu\nu} + h_{\mu\nu}$. With a trivial field space metric we can use the background field approach to obtain Feynman rules with vertex coefficients and propagators given by
\begin{align}
    \lambda_{ab \dots c} =& \i\left\langle\partial_{(a}\partial_b\dots\partial_{c)} S\right\rangle, \\
    \Delta^{ab} =& \i\left\langle\partial_a\partial_b S\right\rangle^{-1},
\end{align}
where $a,b...$ are once again DeWitt indices in either position or momentum space, the factors of $\i$ come from the Wick rotation and $\langle \dots \rangle$ denotes $(\dots)\vert_{\boldsymbol{\Phi} = \boldsymbol{\Phi}_0}$ setting the fields to their background or equilibrium values $\boldsymbol{\Phi}_0$. The $(a,b...c)$ denotes symmeterisation over the indices. To incorporate the field space determinant we can take either of two approaches: either work out the determinant contribution to the effective Lagrangian via 
\begin{align}
    \sqrt{\textup{det}(C_{ab})} &= \exp\left\{\frac{1}{2}\textup{Tr}\left(\ln\left(C_{ab}\right)\right)\right\}, \\
    &= \exp\left\{\frac{1}{2}\int \dd^4 x \sqrt{\bar{g}}\; \textup{Tr}\left(\ln\left(C_{ij}(x)\right)\right)\right\},
\end{align}
then expand in powers of the coupling constants \cite{Herrero-Valea:2020qgj}. Alternatively one can modify the Feynman rules by promoting partial derivatives to covarient field space derivatives \cite{Finn:2021jdz,Finn:2022rlo}
\begin{align}
    \lambda_{ab\dots c} \rightarrow& \i\left\langle\nabla_{(a}\nabla_b\dots\nabla_{c)} S\right\rangle, \\
    \Delta^{ab} \rightarrow& \i\left\langle\nabla_a\nabla_b S\right\rangle^{-1}.
\end{align}
As $S$ is a field space scalar, the $n$-vertex $\lambda_{ab...c}$ is a $\big(\begin{smallmatrix}0 \\ n\end{smallmatrix}\big)$ rank
field space tensor, and the propagator $\Delta^{ab}$ is $\big(\begin{smallmatrix}2 \\ 0\end{smallmatrix}\big)$ tensor. For Feynman diagrams with external legs we also need to define the external factor  
\begin{equation}
    X^a = \left \langle\pdv{\Phi^a}{\chi} \right \rangle,
\end{equation}
where $\chi$ is the physical external field connected to that leg, a field space scalar. This makes $X^a$ a field space vector. The contribution to a matrix element for a particular diagram shape will involve putting combinations of these together, and summing over the field indices. As these are all tensors the resulting object will be a field space scalar, and hence frame invariant. In subsequent sections I will demonstrate how this works for computations of fifth forces, and extend the no-fifth-force result for the scale-invariant theory to all frames.  

\section{Selecting a frame: a geometric approach}

\label{sec:geom}

We can make an observation from section \ref{sec:dilaton}: choosing a frame for a theory with $N$ fields $\{\phi_i,g_{\mu\nu}\}$ is equivalent to taking the $N+1$ field action then imposing a constraint $q(\Tilde{\boldsymbol{\varphi}}) = q(\sigma,\Tilde{\phi}_i) = 0$ on the dilaton and rescaled fields. In terms of the covariant formalism, this means we can consider the field space of the theory in a particular frame as a $N$ dimensional \textit{submanifold} of the appropriate generic $N+1$ dimensional manifold, the field space of a more general theory\footnote{Strictly speaking the field space manifolds are infinite dimensional, as they have $N$ or $N+1$ degrees of freedom at each spatial point. However for clarity I shall just refer to them as ``N" or ``$N+1$ dimensional", counting the number of field species.}. 

Adapting the technique usually applied for gauge fixing, in terms of the path integral we can impose this constraint using delta functions
\begin{align}
    &Z[0] = \int \mathcal{D}^N \Phi \sqrt{\textup{det}(C_{ab})} e^{-S[\boldsymbol{\varphi}]}, \\
    &=\int \mathcal{D}^{N+1} \Tilde{\Phi} \sqrt{\textup{det}(G_{ab})} \prod_x \left[\vert \partial_i q(\Tilde{\boldsymbol{\varphi}}) \vert_x \delta(q(\Tilde{\boldsymbol{\varphi}}(x)))\right] e^{-S[\Tilde{\boldsymbol{\varphi}}]}, \notag
\end{align}
where $G_{ab} = G_{ij}\bar{\delta}^{(4)}(x_i - x_j)$ is the metric on the $N+1$ field space, $\vert \partial_i q(\Tilde{\boldsymbol{\varphi}}) \vert_x$ denotes the magnitude of the field space gradiant of $q$ at space-time location $x$, we count the dilaton $\sigma$ as an additional $\Tilde{\Phi}$ field and we omit the $V_{\textup{gauge}}$ factor for clarity. We may express
\begin{equation}
    G_{ij}\dd\Tilde{\Phi}^i\dd\Tilde{\Phi}^j = \alpha^2 \dd q^2 + 2\beta_i \dd q \dd \Phi^i + C_{ij}\dd \Phi^i \dd \Phi^j.
\end{equation}
where $\alpha = \vert \partial_i q \vert^{-1}$. For $\beta_i = 0$ (we are free to choose this), we have 
\begin{equation}
    G_{ij}\dd\Tilde{\Phi}^i\dd\Tilde{\Phi}^j = \left[\alpha^2 \partial_{\Tilde{\Phi}^i} q \partial_{\Tilde{\Phi}^j} q + \Tilde{C}_{ij}\right]\dd \Tilde{\Phi}^i \dd \Tilde{\Phi}^j.
\end{equation}\\
where $\Tilde{C}_{ij} = C_{kl}\pdv{\Phi^k}{\Tilde{\Phi}^i}\pdv{\Phi^l}{\Tilde{\Phi}^j}$. Then $\textup{det}(G_{ab}) = \textup{det}(\Tilde{C}_{ab}) \prod_x [\alpha(x)^2]$, so 
\begin{equation}
    Z[0] = \int \mathcal{D}^{N+1} \Tilde{\Phi} \sqrt{\textup{det}(\Tilde{C}_{ab})} \prod_x \left[\delta(q(\Tilde{\boldsymbol{\varphi}}(x)))\right] e^{-S[\Tilde{\boldsymbol{\varphi}}]}.
\end{equation}
We may note that $\Tilde{C}_{ab}$ is then the metric one would derive from simply considering the Lagrangian expressed in terms of the $N+1$ $\Tilde{\Phi}^i$ fields. 
We can express the delta functions via the limit 
\begin{widetext}
\begin{align}
    Z[0] &= \lim_{\xi \rightarrow 0} \int \mathcal{D}^{N+1} \Tilde{\Phi} \sqrt{\textup{det}(\Tilde{C}_{ab})} \prod_x \left[\frac{1}{\sqrt{2\pi\xi}}e^{-\frac{1}{2\xi}q(\Tilde{\boldsymbol{\varphi}}(x))^2}\right] e^{-S[\Tilde{\boldsymbol{\varphi}}]}, \\
    &= \lim_{\xi \rightarrow 0} \left[\left(\frac{1}{\sqrt{2\pi\xi}}\right)^{\mathcal{V}} \int \mathcal{D}^{N+1} \Tilde{\Phi} \sqrt{\textup{det}(\Tilde{C}_{ab})} \exp\left\{-\int \dd^4 x \sqrt{-\Tilde{g}}\left(\Tilde{L} - \frac{1}{2\xi}q^2 \right)\right\}\right], \\
\end{align}
\end{widetext}
where $\mathcal{V}$ is an (infinite) measure of the space-time volume. We then have propagator 
\begin{equation}
\begin{split}
    \Delta^{ab} &= \i\left\langle\nabla_a\nabla_b S\right\rangle^{-1}, \\
    &= \i\lim_{\xi \rightarrow 0} \left[\left\langle\Tilde{\nabla}_a\Tilde{\nabla}_b\int \dd^4 x \sqrt{-\Tilde{g}}\left(\Tilde{L} - \frac{1}{2\xi}q^2\right) \right\rangle\right]^{-1}.
\end{split}
\end{equation}
The DeWitt indices are somewhat unwieldy, as they span an infinite number of dimensions. As our theory is local, if DeWitt index $a$ corresponds to position $x$ and field species $i$, then the derivative of the action with respect to $a$ corresponds to a derivative of the Lagrangian at $x$ with respect to field $i$, $\Tilde{\nabla}_a S = \Tilde{\nabla}_i \mathcal{L}\vert_{x}$ where $\Tilde{\nabla}_i$ is the covariant field derivative for metric $\Tilde{C}_{ij}$, allowing us to convert between the two. Assuming a flat background spacetime this gives the finite dimensional propagator
\begin{equation}
    \Delta^{ij} = \lim_{\xi \rightarrow 0} \left[\Tilde{\Delta}^{-1}_{ij} - \frac{1}{\i\xi}q_i q_j\right]^{-1},
\end{equation}
where $\Tilde{\Delta}^{ij}$ is the unconstrained propagator in the $N+1$ field space and $q_i := \left\langle\partial_i q\right\rangle$ the normal covector to the submanifold at $\Tilde{\boldsymbol{\Phi}} = \Tilde{\boldsymbol{\Phi}}_0$. Now consider rotating rotating coordinates in field space such that the direction $q_i$ lies along one axis, call it axis $n$. Then  
\begin{align}
    \Delta^{-1}_{ij} =& \Tilde{\Delta}^{-1}_{ij} - \vert q \vert^2 \frac{\i}{\xi} \delta_{in}\delta_{jn}, \\
    =& \begin{bmatrix}
    \Tilde{\Delta}^{-1}_{pq} & \Tilde{\Delta}^{-1}_{pn} \\
    \Tilde{\Delta}^{-1}_{np} & \Tilde{\Delta}^{-1}_{nn} + \vert q \vert^2 \i \xi^{-1},
    \end{bmatrix}.
\end{align}
where $p,q$ range across all indices other than $n$ and $\vert q \vert$ is the magnitude of $q_i$. Let $\zeta^{-1} = \Tilde{\Delta}^{-1}_{nn} + \vert q \vert^2 \i \xi^{-1}$ (with no summation implied by repeated $n$). Then
\begin{align}
    &\Delta^{ij} = \notag \\
    &\left[\begin{smallmatrix}
    (\Tilde{\Delta}^{-1}_{pq}-\zeta\Tilde{\Delta}^{-1}_{nq}\Tilde{\Delta}^{-1}_{pn})^{-1}  & -\zeta(\Tilde{\Delta}^{-1}_{pq}-\zeta\Tilde{\Delta}^{-1}_{nq}\Tilde{\Delta}^{-1}_{pn})^{-1}\Tilde{\Delta}^{-1}_{pn} \\
    -\zeta\Tilde{\Delta}^{-1}_{nq}(\Tilde{\Delta}^{-1}_{pq}-\zeta\Tilde{\Delta}^{-1}_{nq}\Tilde{\Delta}^{-1}_{pn})^{-1} & \zeta + \zeta^2 \Tilde{\Delta}^{-1}_{nq}(\Tilde{\Delta}^{-1}_{pq}-\zeta\Tilde{\Delta}^{-1}_{nq}\Tilde{\Delta}^{-1}_{pn})^{-1}\Tilde{\Delta}^{-1}_{pn}
    \end{smallmatrix}\right]^{ij}.
\end{align}
To lowest order in $\zeta$ this is 
\begin{align}
    \Delta^{ij} =& 
    \begin{bmatrix}
    (\Tilde{\Delta}^{-1}_{pq})^{-1}  & -\zeta(\Tilde{\Delta}^{-1}_{pq})^{-1}\Tilde{\Delta}^{-1}_{pn} \\
    -\zeta\Tilde{\Delta}^{-1}_{nq}(\Tilde{\Delta}^{-1}_{pq})^{-1} & \zeta
    \end{bmatrix}^{ij}.
\end{align}
where summation is implied over repeated indices $p,q$. We can then see that taking $\xi \rightarrow 0$ and therefore $\zeta \rightarrow 0$ gives 
\begin{equation}
    \Delta^{ij} = 
    \begin{bmatrix}
    (\Tilde{\Delta}^{-1}_{pq})^{-1}  & 0 \;\; \\
    0 & 0 \;\;
    \end{bmatrix}^{ij}.
\end{equation}
This means that taking the limit $\xi \rightarrow 0$ effectively zeros out the contribution from variations in the fields in the direction of $q_i$, which makes sense as we can interpret this as taking the mass of field $q_i \Tilde{\phi}_i$ to infinity. We can write this as 
\begin{equation}
    \Delta^{-1}_{ij} = \left(\delta^k_i - n^k n_i\right)\left(\delta^l_j - n^l n_j\right)\Tilde{\Delta}^{-1}_{kl} = P^k_i P^l_j \Tilde{\Delta}^{-1}_{kl},
\end{equation}
where $P^k_i$ is the projection operator onto the submanifold. This means that we have 
\begin{equation}
    \Delta^{-1}_{ab} = -\i P^c_a P^d_b \langle \Tilde{\nabla}_c \Tilde{\nabla}_d S \rangle = -\i \langle \nabla_a \nabla_b S \rangle,
\end{equation}
where $P^b_a = P^j_i \bar{\delta}^{(4)}(x_a - x_b)$ is the DeWitt-indexed projection operator, $\Tilde{\nabla}_a$ is the covarient derivative on the $N+1$ field manifold, and $\nabla_a$ the covarient derivative on the $N$ field manifold, confirming that this new geometric approach is consistent with the covariant formalism.

\section{Computing fifth forces with the geometric approach}

\label{sec:fith_force}

Let us now see how this geometric approach affects the computation of fifth forces. Our generic Lagrangian is of the form 
\begin{equation}
    \mathcal{L} = \sqrt{-g}\left[F(\boldsymbol{\varphi})R - \frac{1}{2}\partial_{\mu}\phi_i\partial^{\mu}\phi_i - W(\boldsymbol{\varphi}) + L_{\textup{m}} + L_{\textup{gauge}}\right].
\end{equation}
with some integer number of scalar fields. I include a gauge fixing term of the form $\sqrt{-g}L_{\textup{gauge}} = \frac{1}{2}\sigma_{\mu\nu}\Xi^{\mu}\Xi^{\nu}$, which I choose to be the scalar-tensor gauge term used by Copeland et al. \cite{Copeland:2021qby} 
\begin{equation}
    L_{\textup{gauge}} =   \frac{1}{2}F(\boldsymbol{\varphi})g_{\mu\nu}\left[\Gamma^{\mu} - \nabla^{\mu}\ln F\right]\left[\Gamma^{\nu} - \nabla^{\nu}\ln F\right],
\end{equation}
where $\Gamma^{\mu} = g^{\nu\rho}\Gamma^{\mu}_{\nu\rho}$ is the contraction of the spactime connection, $\sigma_{\mu\nu} = F(\boldsymbol{\varphi})g_{\mu\nu}$ and $\Xi^{\mu} = \Gamma^{\mu} - \nabla^{\mu}\ln F$. In place of $V^{-1}_{\textup{gauge}}$ we have the appropriate Fadeev-Popov determinant $V_{FP} = \textup{det}\left(\delta \Xi^{\mu}/\delta \xi^{\nu}\right) := \textup{det}\left(Q^{\mu}_{\nu}\right)$ where $\xi^{\mu}$ are the degrees of gauge freedom \cite{Falls:2018olk,Finn:2022rlo}. This will in turn contribute a ghost term $\mathcal{L}_{gh} = - \bar{c}_{\mu}Q^{\mu}_{\nu} c^{\nu}$ \cite{Falls:2018olk}, but as this only contributes at loop order I shall neglect it here. For the matter terms we include a single fermion field $\psi$, standing in for (eg.) the standard model electron,
\begin{equation}
    L_{\textup{m}} = -\bar{\psi}\left[\i\overleftrightarrow{\slashed{\nabla}} + y(\boldsymbol{\varphi})\right]\psi,
\end{equation}
where the Higgs-like term $\bar{\psi}y(\boldsymbol{\varphi})\psi$ gives the fermion a mass with $y(\boldsymbol{\varphi}) = y_i \phi_i$ (with implied summation). The operator $\overleftrightarrow{\slashed{\nabla}}$ is defined as
$\overleftrightarrow{\slashed{\nabla}} = \frac{1}{2}\left(\overrightarrow{\slashed{\nabla}}-\overleftarrow{\slashed{\nabla}}\right)$
where $\slashed{\nabla} = E^{a\mu}\gamma_a \partial_{\mu}$ is the covariant Dirac operator with $E^{a\mu}$ the vierbein such that $g^{\mu\nu} = E^{\mu a}E^{\nu b}\eta_{ab}$ \cite{Ferreira:2016kxi}. As we now have fermions in our theory, to account for the fermion anticommuntation we need to promote the field space to a \textit{super}manifold \cite{DeWittBryceS1991Ser,RogersAlice2007Ser}, the field space metric to a \textit{super}matrix, and replace the $\textup{det}(C_{ab})$ in the path integral with $\textup{sdet}(C_{ab})$, a \textit{super}determinant. We also redefine $\lambda_{ab\dots c} = i\left\langle\nabla_{\{a}\nabla_b\dots\nabla_{c\}} S\right\rangle$, where $\{a,b,...c\}$ denotes \textit{super}symmeterisation, where we add a factor of $-1$ every time we swap fermion indices. Apart from the need to keep track of minus signs from fermion permutations this does not change the results from the previous sections (more details of the supermanifold construction can be found in \cite{Finn:2022rlo}).

\begin{figure}
    \centering
    \includegraphics[width=0.45\textwidth]{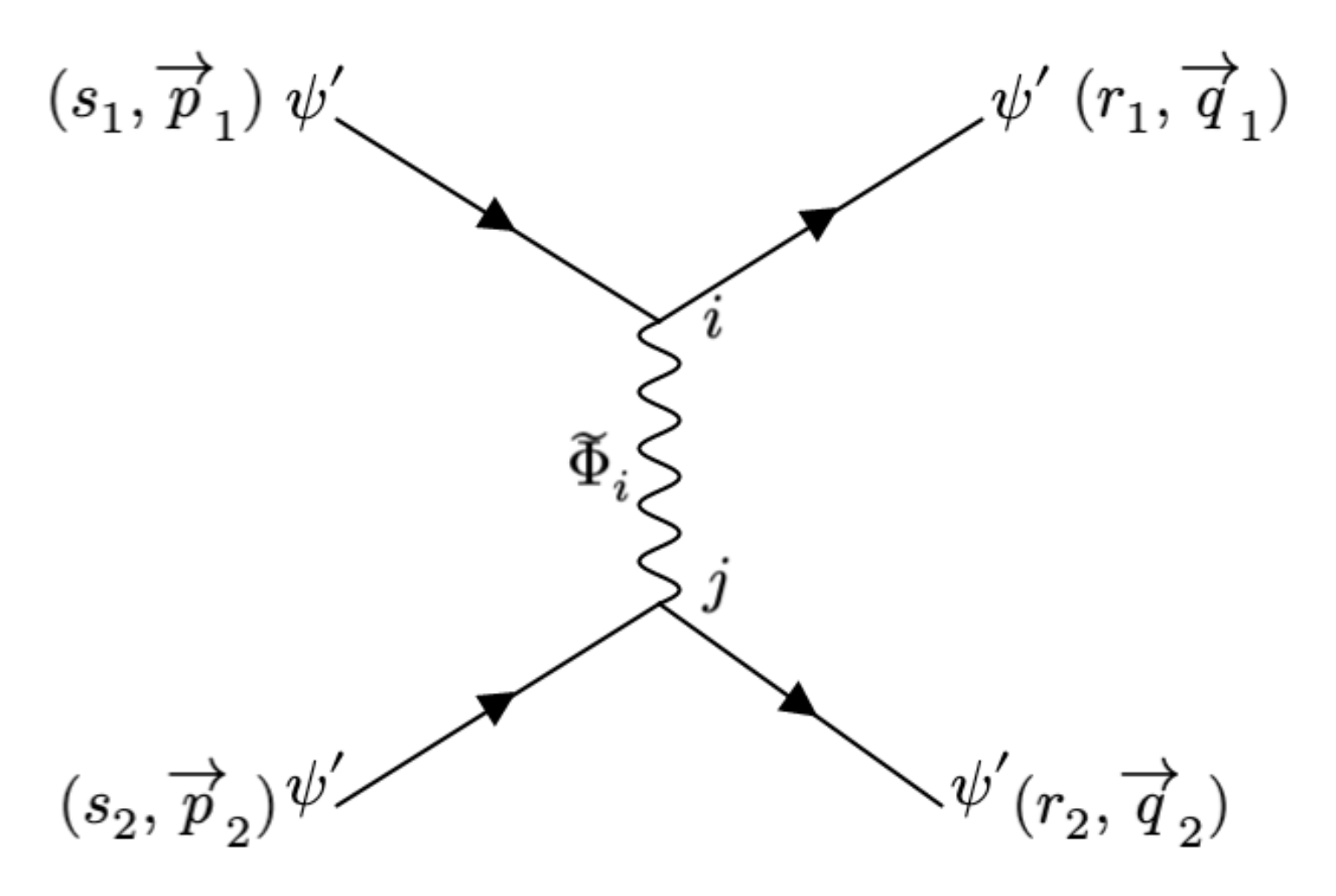}
    \caption{$t$-channel tree diagram for fermion-fermion scattering.}
    \label{fig:tree_diagram}
\end{figure}

Let the total number of field species, including the scalar fields, fermions and graviton, be $N$. Linearising around a background Minkowski metric gives 
\begin{equation}
\begin{split}
    \mathcal{L} =& -\frac{F}{2}\frac{1}{2}P^{\STa\STb,\STc\STd}\partial_{\mu}h_{\STa\STb}\partial^{\nu}h_{\STc\STd} + \frac{1}{2}(\partial_i F)\eta^{\STa\STb}\partial_{\mu}\phi_i\partial^{\mu}h_{\STa\STb} \\
    &- \frac{1}{2}\left(\delta_{ij} - \frac{\partial_i F\partial_j F}{F}\right)\partial_{\mu}\phi_i\partial^{\mu}\phi_j\\
    &- W(\boldsymbol{\varphi}) - \bar{\psi}\left[\i\overleftrightarrow{\slashed{\partial}} + y(\boldsymbol{\varphi})\right]\psi \\
    &+ \frac{1}{2}h_{\STa\STb}P^{\STa\STb,\STc\STd}i\bar{\psi} \gamma_{\STc}\overleftrightarrow{\partial}_{\STd} \psi + \frac{1}{2}h\bar{\psi}y(\phi)\psi + \dots
\end{split}
\end{equation}
where $P^{\STa\STb,\STc\STd} := \frac{1}{2}\left[\eta^{\STa\STc}\eta^{\STb\STd}+\eta^{\STa\STd}\eta^{\STb\STc}-\eta^{\STa\STb}\eta^{\STc\STd}\right]$, $h := h^{\mu}_{\mu}$, here $i,j$ index scalar field species and $\partial_i$ derivatives with respect to the scalar fields, and $\mu,\nu,\STa,\STb,\STc,\STd$ are all spacetime indices. Introducing the general Weyl transformation we obtain\footnote{The ghost terms also rescale such that $\sigma_{\mu\nu} \rightarrow \Tilde{\sigma}_{\mu\nu} = \Tilde{F}(\Tilde{\boldsymbol{\varphi}})\Tilde{g}_{\mu\nu}$ and $\Xi^{\mu} \rightarrow \Tilde{\Xi}^{\mu} = \Tilde{\Gamma}^{\mu} - \Tilde{\nabla}^{\mu}\ln \Tilde{F}$ where $\Tilde{\Gamma}^{\mu}$ and $\Tilde{\nabla}^{\mu}$ are constructed using $\Tilde{g}_{\mu\nu}$} 
\begin{equation}
\begin{split}
    \Tilde{\mathcal{L}} =& -\frac{\Tilde{F}}{2}\frac{1}{2}P^{\STa\STb,\STc\STd}\partial_{\mu}\Tilde{h}_{\STa\STb}\partial^{\mu}\Tilde{h}_{\STc\STd} + \frac{1}{2}(\partial_i \Tilde{F})\eta^{\STa\STb}\partial_{\mu}\Tilde{\phi}_i\partial^{\mu}\Tilde{h}_{\STa\STb} \\ 
    &+ \frac{1}{2}(\partial_{\sigma} \Tilde{F})\eta^{\STa\STb}\partial_{\mu}\sigma\partial^{\mu}\Tilde{h}_{\STa\STb} - \frac{1}{2}\left(\delta_{ij} - \frac{\partial_i \Tilde{F}\partial_j \Tilde{F}}{\Tilde{F}}\right)\partial_{\mu}\Tilde{\phi}_i\partial^{\mu}\Tilde{\phi}_j\\
    &- \Tilde{W}(\Tilde{\boldsymbol{\varphi}},\sigma) + \partial_{\mu}\sigma \left(\partial_i\Tilde{K} - \frac{\partial_{\sigma}\Tilde{F}\partial_i\Tilde{F}}{2\Tilde{F}}\right)\partial^{\mu}\Tilde{\phi}_i \\ &-\left(\Tilde{K}-\partial_{\sigma}\Tilde{K}+\frac{(\partial_{\sigma}\Tilde{F})^2}{2\Tilde{F}}\right)\partial_{\mu}\sigma\partial^{\mu}\sigma \\
    &- \bar{\psi}'\left[\i\overleftrightarrow{\slashed{\partial}} - y_i\Tilde{\phi}_i\right]\psi' + \frac{1}{2}\Tilde{h}_{\STa\STb}P^{\STa\STb,\STc\STd}i\bar{\psi}' \gamma_{\STc}\overleftrightarrow{\partial}_{\STd} \psi' \\
    &+ \frac{1}{2}\Tilde{h}\bar{\psi}'y(\Tilde{\phi})\psi' + \dots \label{eq:Lag_Weyl}
\end{split}
\end{equation}
with an additional field and degree of freedom, and where the fermion field is rescaled as $\psi' = e^{3\sigma/2}\psi$. By adding $\sigma$ we now have $N+1$ field species in the Lagrangian. The equilibrium or background field values define a point in the field space. Expanding around these expected field values, and this point, we have
\begin{equation}
\begin{split}
     \Tilde{\mathcal{L}} &= -\frac{M^2_{\textup{Pl}}}{4}\frac{1}{2}P^{\STa\STb,\STc\STd}\partial_{\mu}\Tilde{h}_{\STa\STb}\partial^{\mu}\Tilde{h}_{\STc\STd} + \frac{1}{2}(\partial_i \Tilde{F})\partial_{\mu}\Tilde{\Phi}^i \partial^{\mu}\Tilde{h}_{\STa\STb} \\
     &- \frac{1}{2}N_{ij}\partial^{\mu}\Tilde{\Phi}^i \partial_{\mu}\Tilde{\Phi}^j - \frac{1}{2}M_{ij} \Tilde{\Phi}^i \Tilde{\Phi}^j \\
     &- \bar{\psi}'\left[\i\overleftrightarrow{\slashed{\partial}} - m_{\Psi} - y_i\Tilde{\Phi}^i\right]\psi' + \frac{1}{2}\Tilde{h}_{\STa\STb}P^{\STa\STb,\STc\STd}i\bar{\psi}' \gamma_{\STc}\overleftrightarrow{\partial}_{\STd} \psi' \\
     &+ \frac{1}{2}\Tilde{h}\bar{\psi}'m_{\psi}\psi' + \dots,  
\end{split}
\end{equation}
where I use $\Tilde{\Phi}^i$ to denote deviations from the background values in all the scalar fields, including the dilaton $\sigma$ and $\Tilde{\phi}_i$. The mass term $M_{ij} = \partial_i \partial_j \Tilde{W}$. We may further simplify this by including the graviton field(s) in $\{\Tilde{\Phi}^i\}$ as well, giving
\begin{equation}
\begin{split}
     \Tilde{\mathcal{L}} &= - \frac{1}{2}N_{ij}\partial^{\mu}\Tilde{\Phi}^i \partial_{\mu}\Tilde{\Phi}^j - \frac{1}{2}M_{ij} \Tilde{\Phi}^i \Tilde{\Phi}^j \\
     &- \bar{\psi}'\left[\i\overleftrightarrow{\slashed{\partial}} - m_{\Psi} - \Tilde{\phi}_i y_i\right]\psi' + \dots,  
\end{split}
\end{equation}
where 
\begin{equation}
    y_{h_{\STa\STb}} = \frac{1}{2}\left[\i P^{\STa\STb,\STc\STd} \gamma_{\STc}\overleftrightarrow{\partial}_{\STd} + \eta^{\STa\STb} m_{\psi}\right].
\end{equation}
and we take the $\overleftrightarrow{\partial}_{\STd}$ operator as only acting on the fermion fields. The lowest order contribution to fermion-fermion scattering is from tree diagrams of the form of Fig. \ref{fig:tree_diagram}. 
This contributes a matrix element of the form
\begin{equation}
\begin{split}
    \i \mathcal{M} =& X^c(\bar{\psi},\vec{p}_1,s_1)\lambda_{cda}X^d(\psi,\vec{q}_1,r_1)\Delta^{ij}\dots\\
    &\dots X^e(\bar{\psi},\vec{p}_2,s_2)\lambda_{efb}X^f(\psi,\vec{q}_2,r_2),
\end{split}
\end{equation}
where here $a,b,c,d,e,f$ are all DeWitt indices. If $\langle \sigma \rangle = 0$, then 
\begin{equation}
    X^a(\bar{\psi},\vec{p}_1,s_1) = \delta^a_{\bar{\psi}'(\vec{p}_1,s_1)}\bar{u}(\vec{p}_1,s_1),
\end{equation}
and 
\begin{equation}
\begin{split}
    &X^c(\bar{\psi},\vec{p}_1,s_1)\lambda_{cda}X^d(\psi,\vec{q}_1,r_1)= \dots\\
    & \dots \i \bar{u}(\vec{p}_1,s_1)\left\langle\partial_{\{\bar{\psi}(\vec{p}_1,s_1)}\partial_{\psi(\vec{q}_1,r_1)\}}\nabla_a S\right\rangle u(\vec{q}_1,r_1), \\
    &= \i \bar{u}(\vec{p}_1,s_1)\left\langle\partial_{\{\bar{\psi}(\vec{p}_1,s_1)}\partial_{\psi(\vec{q}_1,r_1)\}}\partial_a S\right\rangle u(\vec{q}_1,r_1) := \lambda_a.
\end{split}
\end{equation}
Note that this is a field space co-vector. The matrix element is then
\begin{equation}
    \i\mathcal{M} = \lambda_a \Delta^{ab} \lambda_b
\end{equation}
The contribution to the effective potential from this matrix element is given by
\begin{equation}
    V_{\textup{eff}}(r)
    = - \frac{1}{4\pi r} \frac{1}{2 m^2_{\psi}} \sum_{j} \textup{res}_{k=k_j}\left(k e^{\i kr}\mathcal{M}(\boldsymbol{k})\right),
\end{equation}
where $\vec{k} = \vec{p}_1 - \vec{q}_1$ is the exchange momentum between the fermions, and $k_j$ are the poles of the enclosed expression in the upper complex half plane. $V_{\textup{eff}}$ includes both the standard gravitational potential (from graviton exchange) and any fifth force terms.

To evaluate this in a particular frame we need to impose a constraint to go from the $N+1$ fields to the physical $N$ fields. The standard approach is to do this at Lagrangian level. In the Einstein frame $\Tilde{F} = const.$, so if we impose this the explicit coupling between the graviton and the scalar fields dissapears, however there can be a kinetic coupling between the scalar $\Tilde{\phi}_i$ fields and the dilaton due to the 
$(\partial_i \Tilde{K})\partial_{\mu}\Tilde{\phi}_i\partial^{\mu}\sigma$ term. Conversely, if we choose the special Jordan frame where $\Tilde{K} = const.$ then we have a non-trivial kinetic coupling between the graviton and the scalar fields due to $\frac{1}{2}(\partial_i \Tilde{F})\eta^{\STa\STb}\partial_{\mu}\Tilde{\phi}_i\partial^{\mu}\Tilde{h}_{\STa\STb} + \frac{1}{2}(\partial_{\sigma} \Tilde{F})\eta^{\STa\STb}\partial_{\mu}\sigma\partial^{\mu}\Tilde{h}_{\STa\STb}$, 
but remove the kinetic dilaton-$\Tilde{\phi}_i$ couplings. Hence we can think of the frame transformation as a simple exchange between different bosonic degrees of freedom. 

This disadvantage of this approach is that it requires you to redo the entire calculation for each choice of frame, even at tree level, as you need to work out the correct dynamical fields, potential, couplings, and propagator. In the geometric approach from section \ref{sec:geom} we start by calculating everything in the general $N+1$ space. Let $\Tilde{\lambda}_i$ denote the couplings to the external fermion legs in this space for field species $i$, then explicitly we have
\begin{align}
    \Tilde{\lambda}_{\Tilde{\phi}_i} &= \i y_i \bar{u}(\vec{p},s)u(\vec{q},r), \\
    \Tilde{\lambda}_{\sigma} &= 0, \\
    \Tilde{\lambda}_{\Tilde{h}_{\STa\STb}} &= \frac{\i}{4}\bar{u}(\vec{p},s)\left[-P^{\STa\STb,\STc\STd}\gamma_{\STc}(p+q)_{\STd} + 2 m_{\psi} \eta^{\STa\STb}\right]u(\vec{q},r).
\end{align}
The boson propagator with DeWitt indices is given by 
\begin{equation}
\begin{split}
    \Tilde{\Delta}^{ab} &= \i\left\langle\Tilde{\nabla}_a\Tilde{\nabla}_b S\right\rangle^{-1} = \i\left[\left\langle\partial_a\partial_b S\right\rangle + \left\langle\Gamma^c_{ab}\right\rangle\left\langle\partial_c S\right\rangle\right]^{-1}, \\
    &= \i\left\langle\partial_a\partial_b S\right\rangle^{-1},
\end{split}
\end{equation}
as the equilibrium field values satisfy $\partial_a S = 0$, which gives field index propagator 
\begin{equation}
    \Tilde{\Delta}^{ij} = \i\left[-N_{ij} (p-q)^2 - M_{ij}\right]^{-1} = \i\left[N_{ij}t - M_{ij}\right]^{-1}.
\end{equation}
The Mandelstam variable $t = -(q_1 - p_1)^2$. Then to obtain the matrix element in the frame defined by constraint $q(\Tilde{\boldsymbol{\Phi}}) = 0$ we project the tensors onto the submanifold, giving
\begin{equation}
\begin{split}
    \i \mathcal{M} &= (\Tilde{\lambda}_j P^j_i) \lim_{\xi\rightarrow 0}\left[\Tilde{\Delta}^{-1}_{ij} + \frac{i}{\xi}q_iq_j\right]^{-1}(P^k_j \Tilde{\lambda}_k), \\
    &= (\Tilde{\lambda}_j P^j_i)\left[P^l_i \Tilde{\Delta}^{-1}_{lm}P^m_j\right]^{-1}(P^k_j \Tilde{\lambda}_k), \\
    &= (\Tilde{\lambda}_j P^j_i)\left[P^l_i(N_{lm}t - M_{lm})P^m_j\right]^{-1}(P^k_j \Tilde{\lambda}_k).
\end{split}
\end{equation}
All that is needed to evaluate this in different frames is to change the choice of normal vector $q_i$. It may not seem immediately apparent that taking projections of $N+1$ space tensors on different submanifolds should necessarily produce the desired invariant result. However the important point is that only the $N$ field space is really ``physical". The $N+1$ field space is constructed in such a way as to ensure that each submanifold is really just a representation of the $N$ dimensional field space in a different frame, i.e. different field ``coordinates". Hence provided that the underlying objects are tensors, which they are, frame invariance is guaranteed. Also note that at tree level there are no effects from a non-trivial field space metric $\Tilde{C}_{ab}$. These effects can only manifest at 1-loop order or higher. 

\section{Scale invariant theory}

\label{sec:scale_inv}

We can see how this works in the particular case of a scale invariant theory. To make it scale invariant we make $F(\boldsymbol{\varphi})$ quadratic in $\phi_i$ and $W(\boldsymbol{\varphi})$ quartic in $\phi_i$, such that
\begin{align}
    \Tilde{F}(\boldsymbol{\varphi}) :=& -\frac{1}{12}\sum_i \alpha_i {\Tilde{\phi}_i}^2, \\
    \Tilde{K}(\boldsymbol{\varphi}) =& \frac{1}{2}\sum_i (1-\alpha_i) {\Tilde{\phi}_i}^2, \\
    \Tilde{W}(\boldsymbol{\varphi}) :=& \sum_{ij} W_{ij}{\Tilde{\phi}_i}^2\Tilde{\phi}^2_j.
\end{align}
For non-trivial values of $\alpha_i$ the Einstein frame, $\Tilde{F} = const.$, and the particularly interesting Jordan frame, $\Tilde{K} = const.$, describe surfaces, typically ellipsoids, in the $N+1$ dimensional field space. The associated constraints are $q = \Tilde{F} - M^2_{\textup{Pl}}/2$ and $q = \Tilde{K} - K_0$ for the Einstein and special Jordan frame respectively. We can also recover the starting action with the constraint $q = \sigma$, which fixes the dilaton to be zero. For an especially simple example consider a theory with two scalar fields, $\phi_1, \phi_2$ and $\alpha_1 = -1, \alpha_2 = 0, y_1 = 0, y_2 = 1/\sqrt{6}$. In the generalised theory we have three scalar fields plus the graviton $\Tilde{\phi}_i = \{\sigma, \Tilde{\phi}_1, \Tilde{\phi}_2, \Tilde{h}_{\mu\nu}\}$ and 
\begin{align}
    \langle \partial_i \Tilde{F}\rangle &\propto ( 0, \; 1, \; 0, \; 0), \\
    \langle \partial_i \Tilde{K}\rangle &\propto ( 0, \; 2\langle \Tilde{\phi}_1 \rangle, \; \langle \Tilde{\phi}_2 \rangle, \; 0), \\
    \langle \partial_i \sigma \rangle &= ( 1, \; 0, \; 0, \; 0),
\end{align}
as normal vectors to the submanifold for each of the three cases. The fermion mass is $m_{\psi} = \langle \Tilde{\phi}_2 \rangle/\sqrt{6}$ and effective Planck mass $M_{\textup{Pl}} = \langle \Tilde{\phi}_1 \rangle/\sqrt{6}$, which can be used to fix the equilibrium field values $\langle \Tilde{\phi}_i\rangle$ in terms of the masses. The requirement that $\langle\partial_i \Tilde{W}\rangle = 0$ in turn fixes $W_{ij}$ and the mass matrix $M_{ij}$ up to an overall constant\footnote{$M_{ij} = g\left[\begin{smallmatrix} m^2_{\psi} & m_{\psi}M_{\textup{Pl}} \\ m_{\psi}M_{\textup{Pl}} & M^2_{\textup{Pl}} \end{smallmatrix}\right]$ for $i,j$ covering $\{\Tilde{\phi}_1,\Tilde{\phi}_2\}$, for some dimensionless constant $g$.}. Thinking about this theory in terms of the geometric picture outlined in section \ref{sec:geom}, we can immediately see that there is no simple dichotomy between the Einstein frame and the Jordan frame. Instead there is a continuum of frames, characterised by different choices of constraint $q$ and normal vectors $q_i$. Indeed, if we choose a normal vector $q_i = ( \sin \theta, \; \cos \theta, \; 0, \; 0)$ as we rotate in generalised field space from $\theta = 0$ to $\theta = 1$ we can smoothly transform from a completely-Einstein frame to a completely-Jordan frame, encompassing everything in between. Provided one adopts the fully covariant formalism described above, one can be confident of obtaining the same physical results, including the lack of fifth forces, for all frames.  

\section{The field space metric and higher order corrections}

\label{higher_order}

At one-loop order and higher we need to consider the effect of the non-trivial field space metric. One can use the quantum effective action formalism to include quantum corrections non-perturbatively, and much work has gone into developing a frame and/or gauge invariant effective action following the model of DeWitt and Vilkovisky \cite{Vilkovisky:1984st,Christensen:1984dv,DeWitt:1985sg,Ellicott:1987ir,Aashish:2021ero,Giacchini:2020zrl}. Here however I will examine how corrections from the non-trivial metric arise perturbatively purely from the level of the Feynman rules and the geometric approach. As described above, because our theory now includes fermions we need to extend the formalism in section \ref{sec:cov} following the method in Finn et al. \cite{Finn:2022rlo}, promoting the metric to a supermatrix on a supermanifold.
To obtain this metric for our theory we first express the Lagrangian \eqref{eq:Lag_Weyl} as 
\begin{align}
    \mathcal{L} &= - \tfrac{1}{2}N_{AB}(\Tilde{\Phi})\eta^{\mu\nu}\partial_{\mu}\Tilde{\Phi}^A\partial_{\nu}\Tilde{\Phi}^B  - \Tilde{H}(\Tilde{\Phi})^{\mu\nu}i\bar{\psi'}\gamma_{\mu}\overleftrightarrow{\partial}_{\nu}\psi' + \dots,
\end{align}
where $\Tilde{\Phi}^A$ includes all the bosonic fields, this time we have not expanded $N_{AB}$ around the background field values, and $\Tilde{H}^{\mu\nu} = (\eta^{\mu\nu} - \frac{1}{2}\Tilde{h}_{\STa\STb}P^{\STa\STb,\mu\nu} + \dots)$ to first order in fields.
Let $\Tilde{H} := \frac{1}{4}\Tilde{H}^{\mu}_{\mu} = \left(1 + \frac{1}{8}\Tilde{h} + \dots\right)$ then we find
\begin{equation}
    \Tilde{C}_{ij} = -\begin{bmatrix}
    N_{AB} - \frac{\partial_A \Tilde{H} \partial_B \Tilde{H}}{2 \Tilde{H}}\bar{\psi}'\psi' & - \frac{1}{2}\partial_B \Tilde{H} \bar{\psi}' & \frac{1}{2}\partial_B \Tilde{H} \psi' \\
    \frac{1}{2}\partial_A \Tilde{H} \bar{\psi}' & 0 & \Tilde{H} \\
    -\frac{1}{2}\partial_A \Tilde{H} \psi' & -\Tilde{H} & 0 
    \end{bmatrix},
\end{equation}
for fields $\{\Tilde{\Phi}^A,\psi',\bar{\psi}'\}$ and where I have suppressed the fermion spinor indices throughout. The inverse metric is given by 
\begin{widetext}
\begin{equation}
    \Tilde{C}^{ij} = -\begin{bmatrix}
    (N^{-1})^{AB} & - \frac{1}{2\Tilde{H}}(N^{-1})^{AC}\partial_C\Tilde{H} \psi' & - \frac{1}{2\Tilde{H}}(N^{-1})^{AC}\partial_C\Tilde{H} \bar{\psi}' \\
    \frac{1}{2\Tilde{H}}\partial_C \Tilde{H}(N^{-1})^{CB}\psi' & 0 & -\Tilde{H}^{-1} \\
    \frac{1}{2\Tilde{H}}\partial_C \Tilde{H}(N^{-1})^{CB}\bar{\psi} & \Tilde{H}^{-1} & 0 
    \end{bmatrix}.
\end{equation}
The non-zero Christoffel symbols are then
\begin{align}
    \Gamma^{A}_{BC} &= \frac{1}{2}(N^{-1})^{AD}\left(\partial_{B}N_{DC}+\partial_{C}N_{DB}-\partial_{D}N_{BC}\right), \\
    \Gamma^{\psi'}_{BC} &= \left[\frac{\partial_{B}\Tilde{H}\partial_{C}\Tilde{H}}{4\Tilde{H}^2}+\frac{\partial_B\partial_C\Tilde{H}-\frac{1}{2}\partial_D\Tilde{H}(N^{-1})^{DE}(\partial_B N_{EC}+\partial_C N_{EB}-\partial_E N_{BC})}{2\Tilde{H}}\right]\psi', \\
    \Gamma^{\psi'}_{B\psi'} &= \frac{1}{2\Tilde{H}}\partial_B \Tilde{H},
\end{align}
\end{widetext}
plus the appropriate conjugates. However we need to include the correction from the choice of physical space-time metric as discussed in section \ref{sec:cov} giving
\begin{equation}
    \Tilde{C}_{ij}[\textup{correct}] = e^{2(\sigma-\sigma_{\textup{phys}})}\Tilde{C}_{ij}[\textup{from }\Tilde{g}\textup{ frame}].
\end{equation}
Let $\Delta\sigma := \sigma - \sigma_{\textup{phys}}$. If our physical or preferred frame (where $\Tilde{g}_{\mu\nu} = \bar{g}_{\mu\nu}$) is the Einstein frame, then $\Delta\sigma = \ln(2\Tilde{F}/M^2_{\textup{Pl}})$, if it is the $\Tilde{K} = K_0 = const$ Jordan frame we need $\Delta\sigma = \ln(\Tilde{K}/K_0)$, and if it is the original $\sigma = 0$ Jordan frame we have $\Delta\sigma = \sigma$. Note that in all cases $\partial_i \Delta \sigma \propto q_i$. A non-zero $\Delta \sigma$ gives corrections to the field space Christoffel symbols of
\begin{align}
    \delta\Gamma^{a}_{bc} =&  \frac{1}{2}\left(\delta^a_c\partial_{b}\Delta\sigma+\delta^a_b\partial_{c}\Delta \sigma-\Tilde{C}_{bc}\Tilde{C}^{ad}\partial_{d}\Delta \sigma\right),
\end{align}
and thus corrections to the vertex factors $\Tilde{\lambda}_{ab \dots c}$\footnote{If DeWitt index $a$ corresponds to field species $i$ and position $x$ then $\partial_a \Delta \sigma = \partial_i \Delta \sigma \vert_x$.}. Can changing $\Delta \sigma$, and thus the physical spacetime metric, recover a fifth force for the scale-invariant theory? At one-loop order we consider three-point and four point vertices.  For a three-point vertex connected to three bosonic internal fields we have 
\begin{equation}
\begin{split}
    \Tilde{\lambda}_{abc} =& \i\left\langle \nabla_{\{a} \nabla_b \nabla_{c\}} S \right\rangle = \i\left\langle \nabla_{\{a} \nabla_b \partial_{c\}} S \right\rangle \\
     =& \i\left\langle \partial_{\{a} \partial_b \partial_{c\}} S - \left(\Gamma^d_{bc}\partial_a\partial_d S + \Gamma^d_{ab}\partial_c\partial_d S + \Gamma^d_{ac}\partial_b\partial_d S\right) \right\rangle,
\end{split}
\end{equation} so the contribution from $\Delta \sigma$ is  
\begin{equation}
\begin{split}
    &\delta \Tilde{\lambda}_{abc}(\Delta \sigma) = \\ &-\i\frac{1}{2}\left\langle\delta^d_c\partial_{b}\Delta\sigma+\delta^d_b\partial_{c}\Delta \sigma-\Tilde{C}_{bc}\Tilde{C}^{ed}\partial_{e}\Delta \sigma\right\rangle \left\langle \partial_a\partial_d S \right\rangle \\
    &+ \textup{permutations}.
\end{split}
\end{equation}
Let us choose to do the calculation in the preferred frame (remember we are free to choose the frame as the covariant formalism guarantees us frame invariance). Recall that when we apply the frame fixing the propagator will remove any contributions in the direction of $q_i$, so as the terms $\partial_{b}\Delta\sigma$ and $\partial_{c}\Delta \sigma$ only couple to that field direction those terms don't contribute, leaving 
\begin{equation}
\begin{split}
    \delta \Tilde{\lambda}_{abc} =& \i\frac{1}{2}\left\langle\Tilde{C}_{bc}\Tilde{C}^{de}\partial_{e}\Delta \sigma\right\rangle \left\langle \partial_a\partial_d S \right\rangle \\
    &+ \textup{permutations},
\end{split}
\end{equation}
which simplifies to
\begin{equation}
    \delta \Tilde{\lambda}_{abc} = -\i\frac{1}{2}\left\langle N_{bc}M_{ad}(N^{-1})^{de}\partial_e \Delta \sigma \right\rangle + \textup{perms.}
\end{equation}
To get a long-range fifth force potential at loop order we need a $a,b,c$ to correspond to massless fields (the dilaton and the graviton). However if $a,b,c$ are massless, then $M_{ad}=0$ so $\delta \Tilde{\lambda}_{abc} = 0$ regardless of $\Delta \sigma$. 

The other three point vertex we need to consider is one connected to one external fermion leg, one internal fermion field, and one internal massless boson, given by
\begin{equation}
\begin{split}
    \Tilde{\lambda}_{a\psi'}=& \; \i\left\langle \nabla_{\{a} \nabla_{\psi'\}} (X^b\partial_b S) \right\rangle \\
    =& \; \i\left\langle \partial_{\{a} \partial_{\psi'\}} (X^b\partial_b S) \right\rangle - \i\left\langle \Gamma^c_{\{a\psi'\}} X^b(\partial_{c} \partial_b S) \right\rangle,
\end{split}
\end{equation}
where $X^a$ is again the field space vector defining the external, physical, field (note that this means the vertex factor is only a rank-2 field space tensor). Then from the same argument as before we get the contribution from $\Delta \sigma$ as
\begin{equation}
    \delta \Tilde{\lambda}_{a\psi'} = -\i\frac{1}{2}\left\langle\Tilde{C}_{\{a\psi'\}}\Tilde{C}^{cd}\partial_{d}\Delta \sigma\right\rangle  \left\langle X^b(\partial_{c} \partial_b S) \right\rangle.
\end{equation}
For an additional fifth force we need the internal $a$ field to be the dilaton, but $\Tilde{H}$ is independent of $\sigma$ so $\Tilde{C}_{\{\sigma\psi'\}} = \delta \Tilde{\lambda}_{a\psi'} = 0$ regardless of $\Delta \sigma$. While I do not consider the four-point vertex diagrams here, one can in principle calculate corrections from $\Delta \sigma$ for those in a similar manner. The ghost fields do contribute at one-loop order, however by construction the ghost part of the field space metric is trivial, and thus the ghost vertices do not pick up corrections from $\Delta \sigma$ (reflecting the fact they are artificial fields). 

These results suggest that even changing $\Delta \sigma$, which corresponds to changing the \textit{physical spacetime} $\bar{g}_{\mu\nu}$ (or if you prefer Planck length $l(\boldsymbol{\Phi})$), not merely the choice of frame, still does not break the suppression of dilaton fifth forces for scale-invariant gravity, at least to one-loop order. 

\section{Discussion}

In this paper I have shown how one can apply the covariant, geometric formalism to show how scale-invariant scalar-tensor theories evade fifth force constraints in $\textit{all}$ frames. By considering the choice of frame in a fully geometric manner -- as a selection of a submanifold, or normal direction, in the field space of a generalised theory -- we can see that not only is the usual dichotomy of ``Jordan frame" versus ``Einstein frame" really a continuum of frame slices, but that the results of fifth force calculations for any scalar-tensor theory can be made manifestly frame invariant, up to all perturbative orders. Indeed to a large extent we should consider fixing the frame to be directly analogous to fixing the gauge: the frame choice is merely a redundancy of our mathematics\footnote{One could counter by pointing out that, unlike for the gauge, there is always a ``preferred" or ``metric" frame, the one where quantum field $g_{\mu\nu} = \bar{g}_{\mu\nu}$ the metric of physical spacetime. However, absent a full theory of quantum gravity, the extent to which the two should be equal is an open question.}. 

I have neglected vector gauge fields from the matter Lagrangian, however these can be straightforwardly included. A covector field $A_{\mu}$ is Weyl invariant and transforms simply as $A_{\mu} \rightarrow \Tilde{A}_{\mu}$, and the canonical gauge kinetic term $\mathcal{L} \supset -\sqrt{-g}\frac{1}{4}g^{\mu\rho}g^{\nu\sigma}F_{\mu\nu}F_{\rho\sigma}$ is likewise Weyl invariant. Hence adding vector bosons to the scale-invariant theory does not break the scale-invariance, and we conclude that they too decouple from the dilaton \cite{Ferreira:2016kxi}. The addition of vector bosons also does not change our conclusions about general frame invariance: they can simply be included as additional degrees of freedom in our field space, much like the graviton or scalar fields (with appropriate gauge fixing terms). 

While here I have assumed a flat background spacetime, this approach can in be extended to include background spacetimes which are only conformally flat, such as flat FRW. Instead of working on the curved spacetime background, we can change to a frame with a flat background via conformal factor $\Omega = a(\eta)$. The background curvature in one frame can instead be interpreted as a non-zero background value of the dilaton, $\langle \sigma \rangle = \ln(a(\eta))$ in another. Scalar-tensor theories of this form are of cosmological interest as models of inflation \cite{Bezrukov:2007ep, Bezrukov:2012hx, Casas:2017wjh, Hertzberg:2010dc, Ferreira:2016vsc, Hertzberg:2010dc,Karananas:2022byw}, hence it would be worth investigating to see how a similar maximally geometric approach might aid calculations in inflationary background and give confidence when transforming between frames. 

I have also neglected discussion of regularisation and renormalisation, focusing on the results at tree level and lowest perturbative order most relevant for fifth force constraints. Any dimensionful renormalisation scale $\mu$ must transform appropriately between frames \cite{Falls:2018olk}. For the scale invariant theory one can avoid introducing introducing external length scales by making this a function of the scalar fields $\mu = \mu(\boldsymbol{\varphi})$ such that it acquires a stable value in the same manner as the effective Planck mass, and $\mu(\boldsymbol{\varphi})$ then transforms between frames analogous to $F(\boldsymbol{\varphi})$ \cite{Ferreira:2016kxi,Bezrukov:2010jz,Mooij:2018hew,Shaposhnikov:2018nnm,Ferreira:2016wem,Ghilencea:2016ckm,Ghilencea:2015mza}. Usually one would need to specify a frame before renormalising, however it would be interesting to investigate if instead one could first implement a perturbative renormalisation to arbitrary order in the general $N+1$ field theory, as described above, then merely project onto your desired sub-manifold to extract results for a particular frame, and whether one would then naturally obtain the necessary frame dependence for the renormalisation mass scale and other parameters. 

\section*{Acknowledgements}
I thank Pedro G Ferreira for guidance, inspiration and many helpful discussions. I also thank Josu Aurrekoetxea and Katy Clough for helpful advice. This work was supported with funding from a UK Science and Technology Facilities Council (STFC) studentship.

\bibliographystyle{unsrt}
\bibliography{biblio}

\end{document}